\documentclass[]{article}
\usepackage[english]{babel}
\usepackage[utf8]{inputenc}
\usepackage{amsmath}
\usepackage{amssymb}
\usepackage{graphicx}
\usepackage[colorinlistoftodos]{todonotes}
\usepackage{authblk}
\usepackage{hyperref}
\usepackage{ulem}
\usepackage[margin=1in]{geometry}
\usepackage{natbib}
\setcitestyle{authoryear,open={(},close={)}}

\title{Revisiting the Solar Research Cyberinfrastructure Needs: \\A White Paper of Findings and Recommendations}

\author[1]{Gelu Nita}
\author[2]{Azim Ahmadzadeh}
\author[3]{Serena Criscuoli}
\author[3]{Alisdair Davey}
\author[1] {Dale Gary}
\author[4]{Manolis Georgoulis}
\author[5]{Neal Hurlburt}
\author[6]{Irina Kitiashvili}
\author[2]{Dustin Kempton}
\author[1,6]{Alexander Kosovichev}
\author[2]{Piet Martens}
\author[7]{Ryan McGranaghan} 
\author[1]{Vincent Oria}
\author[3]{Kevin Reardon}
\author[2]{Viacheslav Sadykov}
\author[5]{Ryan Timmons}
\author[1]{Haimin Wang}
\author[1]{Jason T. L. Wang}

\affil[1]{New Jersey Institute of Technology}
\affil[2]{Georgia State University}
\affil[3]{National Solar Observatory}
\affil[4]{Academy of Athens}
\affil[5]{Lockheed Martin Advanced Technology Center}
\affil[6]{NASA Ames Research Center}
\affil[7]{Atmosphere and Space Technology Research Associates}

\begin{document}
\maketitle
\begin{abstract}
Solar and Heliosphere physics are areas of remarkable data-driven discoveries. Recent advances in high-cadence, high-resolution multiwavelength observations, growing amounts of data from realistic modeling, and operational needs for uninterrupted science-quality data coverage generate the demand for a solar metadata standardization and overall healthy data infrastructure. This white paper is prepared as an effort of the working group \textit{``Uniform Semantics and Syntax of Solar Observations and Events''} created within the \textit{``Towards Integration of Heliophysics
Data, Modeling, and Analysis Tools''} EarthCube Research Coordination Network (@HDMIEC RCN), with primary objectives to discuss current advances and identify future needs for the solar research cyberinfrastructure. The white paper summarizes presentations and discussions held during the special working group session at the EarthCube Annual Meeting on June 19th, 2020, as well as community contribution gathered during a series of preceding workshops and subsequent RCN working group sessions. The authors provide examples of the current standing of the solar research cyberinfrastructure, and describe the problems related to current data handling approaches. The list of the top-level recommendations agreed by the authors of the current white paper is presented at the beginning of the paper.
\end{abstract}

\section{Executive summary and recommendations}
In this section we summarize the main challenges/issues identified by the authors of this report and the recommendations agreed upon based on their collective expertise in the field without aiming to represent the opinion of the entire community. More detailed supporting arguments and case examples are presented in the subsequent sections.

\subsection{Sustainability and Maintenance of the Heliophysics  Cyberinfrastructure} 
   \begin{itemize} 
    \item \textbf{Challenge:} Currently, the development of data infrastructure projects is incomplete because there is no clear path to fund maintenance or continued development once  the initial project ends.
    \item \textbf{Recommendation:}
    \begin{itemize}
    \item Conduct an evaluation of what it takes to provide sustainable, usable, and interoperable Helio cyberinfrastructure, and of commensurate funding on such projects.
    \item Provide mechanisms for sustainable maintenance of the existing and planned solar data infrastructure elements.
    \end{itemize}
    
    \end{itemize}
\subsection{Storage and Access/Sharing Infrastructure for Heliophysics Data, Models, Analysis and Modeling Tools} 
\begin{itemize}
    \item \textbf{Challenge:} Most of the currently-implemented solutions for data/model storage and sharing are not interoperable. In the advent of an ever increasing volume of pertinent observational data and models, these systems need to be made interoperable.   
    \item \textbf{Recommendations:} 
    \begin{itemize}
        \item Encouragement of uniform semantics and syntax for data services.
        \item Investments into cloud-based storage infrastructure to enable faster data/model transfer.
        \item Investments into cloud-based, on-demand HPC data analysis and modeling infrastructure co-located with the data storage infrastructure.
        \item Encouragement of bringing the diverse community together to work on Helio cyberinfrastructure more cohesively.
    \end{itemize} 
\end{itemize}
\subsection{Data Enhancement}
    \begin{itemize}
        \item \textbf{Challenge:} There are currently limited opportunities for the development of new tools for the Heliophysics data enhancement, even under the existing NASA HDEE program. Smaller communities such as Heliophysics do not see recommendations to apply for funding resources provided by the existing NSF data enhancement programs as viable due to the general nature of the program calls.
        \item \textbf{Recommendation:} Funding opportunities may provide direct references to Heliophysics data enhancement efforts as examples of applicable areas of interest.
    \end{itemize}
 
 \subsection{Discoverability of Existing Heliophysics Data and Tools}

 \subsubsection{Community Awareness on the Existing Resources} 
    \begin{itemize}
    \item\textbf{Challenge:} There is no single place, channel or venue describing the data access and the corresponding software, portals, etc. More organized efforts are needed to address critical questions: Where to access the data? How to access the data? How to post-process the data? How to search for the data?
    \item\textbf{Recommendations:} 
    \begin{itemize}
        \item Support uniform semantics to simplify the search.
        \item {Support the creation of an adaptable, constantly revised online resource describing data access and corresponding software in heliophysics}
        \item Support training workshops and schools for students and postdocs on the available resources, data infrastructure, and metadata/cyberinfrastructure literacy.
  \end{itemize}
  \end{itemize}
 \subsubsection{Automatic Discoverability of Existing Data and Tools}
 \begin{itemize}
     \item \textbf{Challenge:} Most of the solar data and data analysis tools are produced and maintained by instrument teams that do not have significant dedicated support for enabling automatic discoverability. Such efforts require collaboration at a broader level of community involvement.
     \item \textbf{Recommendations:} 
     \begin{itemize}
         \item Create targeted funding opportunities that would encourage and support the practices of using the standardized data models \citep[like SPASE\footnote{\url{https://spase-group.org/}},][]{Roberts2018-SPASE}, standardization of FITS headers, ASDF data format.
         \item Create targeted funding opportunities that would encourage and support the attempts to develop and unify the solar metadata vocabulary and { ontology development for solar events and data}.
         \item{Support the integration and consolidation of the existing infrastructure into a wider scheme}
     \end{itemize}
 \end{itemize}
 
\subsection{Open-Source Software and Rigorous Data Science-for-science Research in Heliophysics}
\begin{itemize}
    \item \textbf{Challenges\footnote{Some of the  recommendations in this section were also included in the \href{https://zenodo.org/record/4048814\#.YUy5geySl7g}{Helio Decadal Survey} white paper by \citet{mcgranaghan2020}}:}
    \begin{itemize}
        \item The traditional incentive in our community is to publish research papers as a primary research outcome and a central metric of the impact and contribution. The Heliophysics community can no longer flourish under these incentive structures alone.
    \end{itemize}
    \item\textbf{Recommendations:}
    \begin{itemize}
        \item Encourage the data science and management components in the same degree as the traditional science components.
        \item Incentivize the development and maintainance of open source software.
        \item Incentives non-traditional research artifacts, such as analysis software, metadata-rich data sets, and collaborative computational resources (such as Jupyter Notebook, JupyterLab, and Google Colaboratory).
        \item Increase support for Heliophysics data infrastructures that serve AI/ML such as analysis-ready data and pipelines to create them.
        \item{Invest in  interpretable machine learning to enhance heliophysics in par with data science developments in a research-to-operations-to-research (R2O2R) feedback loop.}
    \end{itemize}
\end{itemize}

\section{EarthCube Research Coordination Network:  Towards Integration of Heliophysics Data, Modeling, and Analysis Tools (@HDMIEC RCN)}
    The authors of this report are members of a Research Coordination Network (RCN) that is supported by NSF through the EarthCube (EC) AGS-1743321 grant to the New Jersey Institute of Technology. Any opinions, findings, conclusions, and recommendations expressed in this text are those of the authors and do not necessarily reflect the views of the National Science Foundation. This EC RCN, named ``Towards \textbf{I}ntegration of \textbf{H}eliophysics \textbf{D}ata, \textbf{M}odeling, and \textbf{A}nalysis \textbf{T}ools" (\textbf{@HDMIEC}), was created to foster new collaborations between heliophysics and computer science research communities. 
    
    To make progress in this front, the heliophysics research community needs to (i) understand the physical mechanisms behind the observed complex, multi-scale processes on the Sun and (ii) develop practical and accessible cyberinfrastructures aiming to collect, access, analyze, share, and visualize all forms of data from ground- and space-based observatories, as well as data-driven numerical simulations and modeling. These needs and goals require a synergy between classical solar physics, data science, and computer science. The @HDMIEC RCN actively encourages the involvement of a diverse pool of community members and provides an organized virtual structure that promotes collaboration on open-source knowledge discovery. This collaboration aims to enhance the science return of individual research projects run by the RCN participants. 
    
    \subsection{Activities Leading to the Current Report}
        Up to date, the @HDMIEC RCN has been conducting its activities in the real or virtual space through a series of monthly working group virtual meetings focused on different topics of interest, as well three community-wide workshops organized by its steering committee.
    
    \subsection{Kickoff Workshop}
        The first @HDMIEC RCN workshop was held on 14-16 November 2018 at NJIT, was attended by 43 participants, and featured 19 invited presentations, including 7 presentations by representatives of other domains of geophysics funded by the EarthCube program. The main goal of this workshop was to reiterate and discuss the main challenges faced by the heliophysics community from the perspective of its needs for more efficient integration of data, modeling, and analysis tools, and to organize its future activities aiming to address these challenges. 
    
    \subsection{Working Group Workshop and White Paper: {\it``Machine Learning in Heliophysics and Space-Weather Forecasting"}}
        A second @HDMIEC RCN workshop was held on 16-17 January 2020 at NJIT. This workshop brought together a group of 40+ heliophysicists, data providers, expert modelers, and computer/data scientists. Their objective was to discuss critical developments and prospects of the application of machine and/or deep learning techniques for data analysis, modeling, and forecasting in Heliophysics, and to shape a strategy for further developments in the field. The workshop combined a set of plenary sessions featuring invited introductory talks interleaved with a set of open discussion sessions. The outcome of this workshop was presented in a white paper that featured a top-level list of recommendations \citep{RCN2020ML}
    
    \subsection{Working Group White Paper: {\it ``Challenges and Advances in Modeling of the Solar Atmosphere: A White Paper of Findings and Recommendations"}}
        Another set of challenges faced by the heliophysics research community was addressed by \citet{RCN2021WP} in a white paper that was the outcome of one of its focused working groups, which met regularly in the virtual space.
    
    \subsection{Working Group Virtual Workshop: {\it``Uniform Semantics and Syntax of Solar Observations and Events"}}
        Another activity organized by the @HDMIEC RCN was a special working group session at the EarthCube Annual Meeting on June 19, 2020, which has gathered 20+ virtual participants. This white paper summarizes the  presentations and discussions held during this virtual workshop, as well as contributions from a wider community, which were gathered during a series of preceding and subsequent working group sessions.
        
    \subsection{Placing the Findings and Recommendations of This White Paper in a Historical Perspective }
        The authors of this white paper also attempt to place their findings and recommendations in a historical context by taking advantage of the fact that, although the activities of this particular RCN have only a three-year span, most of its members participated in several preceding activities spanning a much longer period,  which motivated and helped the creation of this RCN. 
        
        The most recent such seeding activity was the workshop entitled {\it ``Roadmap for Reliable Ensemble Forecasting of the Sun-Earth System,"} which was held 28-30  March  2018 at NJIT and was attended by 47 participants representing 25 institutions, who contributed to a white paper presenting their findings and recommendations \citep{Roadmap2019}.
        
        Nevertheless, the first of these seeding activities was the EarthCube End-User Workshop held on 13-15 August 2014 at NJIT entitled {\it``Science-Driven Cyberinfrastructure Needs in Solar-Terrestrial Research,"} which gathered 80 solar-terrestrial and computer science researchers. Given the importance of the findings and recommendations resulted from this workshop that were submitted in the form of an Executive Summary to NSF, but never published in a permanent repository, we include, for completeness and future reference, a copy of this summary in Appendix \ref{STR} to this white paper.

\section{Current Standing of Solar Research Cyberinfrastructure}
    
    \subsection{Diversity and Complexity of the Solar Data and Metadata}
    
        Advances in heliophysics are, to a significant extent, driven by the analysis and interpretation of observational data. Starting from the launch of the Solar Dynamics Observatory in 2010, solar observations resulted in more than 2\,TB of data volume per day. Nowadays, more than 8-10\,TB daily volume rates are expected from the Daniel K. Inouye Solar Telescope (DKIST) when routine operations commence. As a result, the total volume of the accumulated solar observational data is already around 10\,PB and is projected to increase by about 4\,PB per year. To efficiently process and analyze this amount of data and also integrate them into event catalogs, uniform metadata definitions and structures are needed. At the same time, the development of encompassing, comprehensive and, importantly, user-friendly cyberinfrastructure is essential. What complicates this challenge is that solar observations themselves vary in nature, from Sun-as-a-star observations of integrated intensities in particular wavelength ranges, to essentially four-dimensional imaging spectroscopy scans of selected parts of the solar disk. One may also add here data coming from various attempts to model the solar dynamics at different scales, and receive a complicated and non-easily discoverable data system as a result.
        
        In addition to the work and challenges related to observational data processing and provision, there are numerous attempts targeted on recognition of features in solar data (e.g. solar flares, sunspots, coronal holes, filaments, etc), quantification of these features in terms of their properties, and association of features with each other (for example, linking of the flare to a host solar active region). As an example, the Heliophysics Event Knowledgebase (HEK) collects more than 25 classes of features relying on data from different instruments. Some of these features are identified from time series analysis (e.g. soft X-ray solar flares, solar energetic particle events, etc), while others are results of time-dependent image processing (detected active regions, sunspots, filaments, etc.) or processing of solar spectra (e.g. Ellerman bombs or UV bursts). Moreover, particular classes of events (e.g., solar flares) are reported from data of different telescopes with these reports not necessarily integrated with each other. Yet, metadata supporting the features allows to associate them with observational data and encourages statistical studies.
            
        Given this variety of data and derived data products/features, the questions of uniform semantic and syntax, as well as the development of the approaches enhancing the data availability, discoverability, and scientific readiness, become especially important. In the following sections, we provide \textit{high-level example-based descriptions} of the current massive data / metadata / data handling approaches, discoverability attempts, as well as outstanding problems in these topics.
    
    \subsection{Current Massive Data, Metadata, and Data Product Handling Approaches\label{data_volume}}
    
    \subsubsection{Data and Products Maintained at Observing Facilities}
        
        The large data volume of spectral and spectro-polarimetric observations of the Sun, together with embargoes often imposed on the data, have long posed serious limitations on the development of solar observation databases. Recent attempts have been so far mostly limited to efforts of individual instrument teams or specific projects. Recent efforts include: data acquired during different service mode campaigns at the Dunn Solar Telescope (DST) \footnote{ https://nso.edu/telescopes/dunn-solar-telescope/dst-smo-2/}; a database of observations acquired with the Interferometric Bidimensional Spectrometer at the DST \footnote{http://ibis.oa-roma.inaf.it/IBISA/}; observations acquired with the Goode Solar Telescope (GST) \footnote{http://www.bbso.njit.edu/~vayur/NST\_catalog/}; a database of observations acquired with the Fast Imaging Solar Spectrograph at GST \footnote{http://fiss.snu.ac.kr/data\_catalog/}; a set of IRIS and Swedish Solar Tower coordinated observations \footnote{https://iris.lmsal.com/search/}; the database of observations acquired with the Expanded Owens Valley Solar Array (EOVSA) \footnote{http://ovsa.njit.edu/browser}. The majority of these databases have been developed in the recent past thanks to the increase of internet bandwidth and disk storage. 

    \subsubsection*{Big Bear Solar Observatory}
    
        NJIT has built, and now operates the second largest aperture ground-based solar telescope in the world- the 1.6-meter Goode Solar Telescope \citep[GST,][]{GST} at the Big Bear Solar Observatory (BBSO), California.  Benefiting from the long periods of excellent local seeing at the Big Bear Lake, the GST, equipped with high-order adaptive optics, routinely collects diffraction-limited spatial resolution (0.1") photometric, spectroscopic and/or polarimetric data, with a high cadence (40 s), across the spectrum from 0.4-5.0 $\mu$m. Since the beginning of its regular operation in 2010, it has provided the community with open access to observations of the photosphere, chromosphere and up to the base of the corona with unprecedented resolution, targeting at the fundamental nature of solar activity and the origin of space weather. 
    
        In more detail, the Visible Imaging Spectrometer (VIS) and a high-cadence, large-format imager acquire, for example, fast scans of the H$\alpha$ line and TiO filtergrams to monitor chromospheric dynamics and small-scale (magnetic) brightenings, respectively. Both types of data benefit from post-facto image restoration, thus reaching the telescope's diffraction limit. The most important development is the Near InfraRed Imaging Spectropolarimeter (NIRIS), which has a high Zeeman sensitivity and profits from the improved seeing conditions in the NIR. The primary line of interest for NIRIS is the Fe\,\textsc{i} 1565~nm doublet, which is the most Zeeman sensitive probe of the magnetic field in the deepest photosphere. The volume of raw data is about 4TB/day. The processed data is around 100GB/day,  which can be downloaded from a data portal at BBSO. 

    \subsubsection*{Expanded Owens Valley Solar Array}
    
        The Expanded Owens Valley Solar Array (EOVSA) was completed in 2017, late in the previous solar cycle \citep{Gary_2018}.  The 13-antenna array has recently (2021) been designated an NSF Geospace Facility.  It observes the entire Sun daily in the range 1-18 GHz with the ability to create multi-frequency images at up to 451 frequencies per second.  Although the raw data volume is relatively modest by today's standards (50 GB/day), the creation of images from the data, spectral analysis, and modeling may imply more than $\sim15\times$ growth in data volume, which represents a considerable challenge in data processing, storage, and dissemination.

        Currently the raw data are processed by a pipeline that produces daily full-disk images at 7 frequencies, at the same time reformatting the data for the CASA (Common Astronomy Software Applications) standard.  As solar activity increases, a second radio flare pipeline is being developed to provide standard image sequences during flares with 10 s cadence at 50 frequencies. Individual events can also be processed fully at 1 s cadence and at even higher frequency resolution if warranted by the spectral and dynamic complexity. These flare datacubes can then be processed with gyrosynchrotron spectral fitting to create movies of parameter maps (e.g. coronal magnetic field strength, nonthermal electron number density and powerlaw index, and thermal temperature and density). The final step in flare analysis is the creation of 3D models of the sources that fit the imaging and spectral constraints not only from microwaves, but including EUV, X-rays, and other data.

        The raw data and pipeline data products are available from a web portal in standard formats organized according to time, but at present there is no system for searching the database or metadata on other parameters.  Adding products from the associated analysis and modeling results in the form of a searchable database remains unexplored, yet would clearly be highly advantageous to the science community.
        
    \subsubsection*{Daniel K. Inouye Solar Telescope Data Center}
    
        The US National Science Foundation’s (NSF) Daniel K. Inouye Solar Telescope (DKIST) \citep{Rimmele2020} is a ground based solar telescope located on Haleakalā, Maui, Hawai’i. The telescope is currently the largest optical solar telescope in the world combining a 4-meter primary mirror with a ground-breaking 5 instrument suite. DKIST delivers spatial resolution and sensitivity that enable solar physicists to unravel many of the mysteries that the Sun presents, including the origin of solar magnetism, the mechanisms of coronal heating and drivers of the solar wind, flares, coronal mass ejections, and variability in solar output.

        DKIST will deliver approximately 3 PB of raw data per year. Calibrated data volumes will be of similar magnitude. These data as well as higher-level data products have to be curated over the lifetime of the observatory, two full Hale cycles, or approximately 44 years.
    
        The DKIST Data Center represents a major transition in how users are granted access to facility data. In contrast to the previous national ground-based solar facilities, the DKIST program includes a data center that will provide calibrated data to users. By making calibrated and eventually higher-level data products broadly available, the scientific utility and impact of DKIST is greatly enhanced. The Data Center’s mission includes:
    
        \begin{itemize}
            \item Delivering quality-controlled, calibrated scientific data sets from data acquired by DKIST facility instruments utilized during the operational lifetime of DKIST.
            \item Providing long-term management and curation of DKIST scientific data and metadata resulting from DKIST experiments.
            \item Enabling open, searchable, and documented access to data and metadata to a broad user base.
            \item Developing and maintaining software and hardware tools to support the acceptance, usage, and exploitation of DKIST data, accommodating changing scientific and operational needs inherent in a multi-decadal scientific program.
        \end{itemize}
        The Data Center provides the crucial link between the science data and associated metadata acquired at the telescope and enabling DKIST scientific goals. DKIST will deliver approximately 8-10 TB of data daily, that will be calibrated to remove telescope, instrumental and atmospheric seeing effects. Dealing with this kind of data flow requires automated calibration which in turn requires automated quality control to ensure the data that is being supplied to the community is of the highest quality. Furthermore, the distribution of large DKIST data sets in an efficient and reliable manner to a widely distributed user base with the required user support, presents significant technology challenges and operational effort. In addition, the Data Center must be a flexible system that accommodates scientific, instrument and technological changes. This has to be done while maintaining operational status and data and processing provenance. 
        
        The Data Center has been designed around a microservice event driven architecture which allows the deployment of small changes easily, without having to disrupt the entire system with large scale outages and will allow the Data Center to easily deploy additional resources where they are required. The Data Center is being developed using Continuous Integration / Continuous Delivery methodologies with a strong focus on testing at all levels. The goal is for all tests to be automated instead of manual to help speed up deployment cycles. This is helping to enable a small team to develop a Data Center capable of dealing with the technical and scientific challenges posed by DKIST. 
        
        The DKIST DC will enable open, searchable, and documented access to data and metadata by a broad user base, as well as providing python-based User Tools to enhance these processes and scientific interaction with the data.
    
    \subsubsection*{Global Oscillation Network Group}  

        Another successful attempt towards increasing the continuity and accessibility of data is the Global Oscillation Network Group \citep[GONG,][]{1997SPD....28.0211L} which is a network of six instruments, namely Learmonth (Australia), Udaipur (India), El Teide (Spain), Cerro Tololo (Chile), Big Bear (USA), and Mauna Loa (USA). The full network of the GONG became operational in January 1995, with low-resolution cameras ($204\times239$ pixels). After 3 years of operation, it was upgraded to $1024\times1024$ pixels (commonly referred to as the GONG+). In the following years, the GONG's capabilities were extended to address the space-weather research needs as well. In 2006, its resolution was upgraded again to $2048\times2048$ pixels, and its cadence, to less than one minute per observation. The integration of these six instruments resulted in the mean duty cycle (i.e., the fraction of 24-hour period observations are available) of 93\% for the last 18 years of its (25-year) ongoing operation \citep{2021PASP..133j5001J}. The GONG provides two main products, namely the Full-Calibration and Near-Real Time products. The former includes (1) Magnetogram, Velocity \& Intensity, (2) Global Helioseismology, (3) Local Helioseismology, and (4) Magnetic-Field Products. The latter includes (1) Magnetogram \& Intensity, (2) H-Alpha, (3) Farside Images, (4) Corrected Magnetic-Field Products, and (5) Uncorrected Magnetic-Field Products. All observations made by the GONG network of instruments are publicly available through a web interface (as well as an FTP archive) that provides filters for queries of specific products, instruments, time interval, etc.
    
    ~
    
    \noindent Although maintaining the databases of observational data and products at observing facilities is undeniably valuable, which constitutes a big improvement with respect to the past, there are still several limitations of that approach. For instance, most of these databases have limited (if any) search capabilities; data acquired with the same instruments or of the same target are often scattered in different databases; data acquired with the same instrument but available through different institutes may be reduced with different software packages, which might introduce inconsistencies difficult to asses by non-expert users.
    
    \subsubsection{Metadata Collection and Centralization Attempts}

    \subsubsection*{The Virtual Solar Observatory}
        The Virtual Solar Observatory \citep[VSO\footnote{https://virtualsolar.org},][]{Hill2009}  is a project designed to help the Solar and Heliophysics community, by providing a resources that allow users to discover, search filter and download data sets they need to do their science. By providing a unified interface for these tasks the VSO has streamlined the steps required to get data from the archives to the user. 
        The VSO design goals were:
        \begin{itemize}
            \item Provide unified access to searches performed on distributed data archives. 
            \item Provide uniform interface for searches and data requests.
            \item Allow multiple data servers and providers.
            \item Allow multiple user and application interfaces  
            \item Require minimal effort for provider participation, and make best use of existing services.
            \item Allow for minimal centralization of metadata.
        \end{itemize}
        VSO currently supports accessing data from 28 different sites, covering 51 different sources and 103 different instruments.\footnote{https://sdac.virtualsolar.org/cgi/show\_details?keyword=INSTRUMENT} The available data includes space, ground-based, rocket flight data and historic data going all the way back until 1869\footnote{http://mlso.hao.ucar.edu/hao-eclipse-archive.php - High Altitude Observatory Eclipse Archive}. The VSO provides a standard web interface by which users can perform data searches. Through its application interfaces, the VSO is tightly coupled to both major solar physics analysis environments, Solarsoft\footnote{https://www.lmsal.com/solarsoft/} and SunPy\footnote{https://sunpy.org/}. In both environments a user can search, filter and download data without having to leave the environment, and then move straight to calibration and analysis.
    
        In its support of the Solar and Heliospheric physics community VSO has sought to improve the state of data archiving by, for example;
        \begin{itemize}
            \item Providing data validation services.
            \item Developing systems, that adaptively include existing data  sets.
            \item Providing a simple and easy path for the addition of new sets.
            \item Developing best practices for creating data archives and research catalogs.
            \item Facilitating data mining and content-based data searches.
        \end{itemize}

    \subsubsection*{Heliophysics Event Knowledgebase} 
        {Among the attempts to capture, share, and annotate solar and space weather phenomena and data, Heliophysics Event Knowledgebase \citep[HEK,][]{Hurlburt2012hek} is one of the most notable. HEK is an event-centric database currently containing 24 event classes (e.g. solar flares, emerging flux events, filament eruptions, etc) resulting in more than 1.7 million individual event records. HEK also provides the instrumental coverage information (currently 10 instruments from 5 missions, resulting in 160 thousand coverage records). The events are determined empirically, based on where, by what algorithm, and by using what data the event was found. HEK provides a high degree of flexibility and interaction with end users by a REST API, allowing to search and request the records and supporting the events and coverage records by a machine-readable standardized metadata. The capabilities to explore the HEK are integrated with the packages widely adopted by the solar community, such as IDL SolarSoft and Python SunPy. In addition to queries via API, HEK provides a web interface for the search, quicklook information and related visualizations/data for some events. A knowledgebase is a continuously-developing product: some of the near-term plans include a support for the recent off-axis solar missions (such as the Parker Solar Probe and Solar Orbiter), ground-based telescope data, prediction and modeling attempt results, as well as transitioning to more community-supporting model operation.}

\subsection{Data Discoverability and Enhancement Services}
    
    \subsubsection{Feature Recognition and Finding Attempts}
    
    \subsubsection*{Feature Recognition, Feature Finding Modules at HEK}

The Solar Dynamics Observatory (SDO) produces about 4k $\times$ 4k 100,000 images per day, far more than any previous solar physics mission. NASA recognized well before the SDO launch that this torrent of data would not be easily digestable by solar physicists using the same methods that previous missions have employed. In order for the community to use the SDO data effectively the SDO Feature Finding Team (FFT) \cite{martens-etal-2012} started producing, well before launch, a set of robust and efficient professionally coded software modules capable of keeping up with the relentless SDO data stream.  These modules detect, trace, and analyze a large number of phenomena, including: flares, sigmoids, filaments, coronal dimmings, polarity inversion lines, sunspots, X-ray bright points, active regions, coronal holes, EIT waves, CMEs, coronal oscillations, and jets. Other modules track the emergence and evolution of magnetic elements. The detection of CMEs and filaments is accomplished with SoHO/LASCO and ground based H$\alpha$ observations.

The metadata produced by the FFT module are delivered mostly to the Heliophysics Event Knowledge base (HEK) \citep{Hurlburt2012hek}, and also to various catalogs for CMEs and filaments. These metadata allow solar physicists to identify and retrieve data sets relevant to their research projects.  The FFT has achieved the first comprehensive computer vision post-processing pipeline for solar physics, abstracting complete metadata on solar features and events without human
intervention.

Since the completion of the FFT project in 2012 great strides forward have been made in feature recognition and tracking in solar physics through the application of modern Data Mining (DM) and Machine Learning (ML) techniques.  Instead of the development of modules specifically targeted at one solar feature (say H$\alpha$ filaments), multi-purpose ML modules have been developed, trained by a set of images of a targeted feature, and then detect the same feature in any other images produced by the same instrument.  The same ML module can then be employed for detection of other features.  The first such module for solar physics is described in \cite{banda-etal-2013}.  ML-ready online datasets have been created from these metadata \citep{galvez-etal-2019, angryk2020multivariate}, and a multi-purpose tracking module for solar imagery has been developed by \cite{berkay-angryk-2018}. Current and future research in solar physics will greatly benefit from this and similar new tools and repositories.

    \subsubsection*{Feature Finding -- GSU Data Mining Lab Efforts}
    
        Towards creation of an Content-Based Image Retrieval (CBIR) system, DMLab has conducted several studies in the past decade \citep[e.g.,][]{kempton2016towards, schuh2017region, ahmadzadeh2017improving}, which resulted in the release of a large dataset of image parameters extracted from SDO mission's AIA instrument. This dataset provides ten pre-computed image parameters on the nine wavelength channels of the instrument, covering AIA images since 2011 (and counting), with the cadence of 6 minutes. This 1TiB-per-year dataset provides a much lighter version of the SDO's 0.55 petabytes per year \citep{martens-etal-2012}, tailored for tasks related to CBIR, including feature recognition. Tracking solar events is one of the direct contributions of this dataset \citep{kempton2018tracking}. A thorough discussion on the collection, curation, and integration of this dataset, as well as the data analytics and feature recognition experiments are authored by \cite{ahmadzadeh2019curated}, and the project is made open-sourced as part of the \cite{imageparameter2019bitbucketurl}. The dataset is locally maintained on the DMLab's server and is publicly available via a REST API \citep{imageparameter2019api}.

    \subsubsection{Data and Metadata Integration and Visualization}
    
    \subsubsection*{NASA NAS Helio Portal}
        The solar data are often stored at individual locations and catalogs and do not often interact with each other. One of the possibilities to enhance solar data exploration and discoverability and foster multi-messenger data analysis is to integrate the various sources and provide a convenient interface to explore the integration results. The examples of the data integration attempts are the services at the NASA Heliophysics Modeling and Simulations Data Portal (HMS portal, \url{https://data.nas.nasa.gov/helio/}). The Interactive Multi-Instrument Database of Solar Flares \citep[IMIDSF,][]{Sadykov2017-IMIDSF} combines the flare records from numerous sources (e.g., GOES flare catalog, RHESSI flare catalog, HEK ultraviolet flare records, etc.). The key capabilities of the database are the search for the flare events simultaneously observed by several instruments and quicklook visualization tools, which simplify the search and preparation of events of interest for the statistical and case studies. Another example is the Radiation Data Portal \citep[RDP,][]{Sadykov2021-RDP} also located at the HMS portal. The portal connects the radiation measurements obtained from the Automated Radiation Measurements for Aerospace Safety (ARMAS) device and the soft X-ray and proton flux measurements from GOES, allowing one to explore the solar activity during the flight. Both portals are deployed to NASA, however, both require constant attention and modifications, raising a sustainability question sharp. In addition to two portals mentioned above, the NASA Ames Helio Portal provides access to the heliophysics modeling data sets of global and local simulations of the Sun and corresponding observables.
        
    \subsubsection*{Convergence Hub for the Exploration of the Space Science}
    
        The operational demands of space weather coupled with the scientific needs of Heliophysics demands a new approach, one that integrates innovative ideas, approaches, and technologies from across the entire Sun-to-Earth Infrastructure. Since 2019, the Convergence Hub for the Exploration of Space Science (CHESS) project has been pioneering this ``convergent'' approach, focusing on the space weather impact to the power grid as a vital and demonstrable use case. The CHESS vision is to combine and visualize datasets together with the application of advanced data science and machine learning capabilities into a novel networked computational infrastructure, enabling both power utility users and space weather researchers to engage with the data in meaningful ways: an Open Knowledge Network. The domain and cyberinfrastructure that CHESS created converges the domains involved (from solar and space physics, to geoscience, electrical engineering, and computer and data science) and produces the advanced computational tools to quantify and protect against the risks of space weather-induced power outages. The particular implementation steps of CHESS include (1) a use case/user-centered design approach to highlight the relevant space weather information to the electric power grid operators based on their needs, (2) a new process for working with ontology engineers following the modular ontology design principles, (3) a transition from ontology design patterns to build an open knowledge network, and (4) a creation of dashboards that allow users to interact with the network and extract/visualize necessary information.

        The CHESS National Science Foundation Convergence Accelerator project designed tools to connect disparate data sources via a knowledge network. It set a precedent for domain and ontological scientists to co-develop an ontology (and the technology stack required), and showcased the importance of a user-centered design approach and working with designers to make the product usable by different communities.

\subsection{Integration of the Solar Data with the Modeling Efforts}

    \subsubsection{Data-constrained Modeling Efforts}
        Modern Geospace science explores the Sun and its effects on Earth through new observations with substantially improved spatial, spectral, and temporal resolution relative to only a few years ago, brought about by recent advances in technology.  To make sense of these highly demanding data, heliophysicists develop sophisticated models that integrate the most advanced theories, computational codes, and simulation tools. A great science challenge is to tightly couple the observations and theory-driven modeling to obtain fundamental knowledge that can lead to a deeper understanding of the mechanisms of solar variability as well as to new physics-based approaches to predicting solar magnetic activity and solar storms. A key component in addressing this challenge is the understanding of the 3D Structure of the solar atmosphere and magnetic field through realistic modeling simultaneously constrained by the available multiwavelength observations \citep{RCN2021WP, Gibson2020}.  

        An example of a community-wide data-to-model integration effort  is the fully automatic model production pipeline (AMPP), which is meant to facilitate the use of the GX\_Simulator Solar Soft/IDL data-contrained 3D flare and Active Region modeling package \citep{Nita2015,Nita2018}.
        Based on minimal users input, AMPP downloads the required vector magnetic field data produced by the Helioseismic and Magnetic Imager (HMI) onboard the Solar Dynamics Observatory \citep[SDO,][]{Scherrer2012} and performs potential  and/or nonlinear force free field (NLFFF) extrapolations. Optionally,  AMMP also downloads the contextual Atmospheric Imaging Assembly maps \citep[AIA,][]{Lemen2012}, which may be used for model constraining through model-to-data comparison. Then, AMPP populates the magnetic field skeleton with parametrized heated plasma coronal models that assume either steady-state or impulsive plasma heating, and generates non-LTE density and temperature distribution models of the chromosphere that are constrained by photosphere-level SDO/HMI measurements. The standardized models produced by this pipeline may be further customized through a set of interactive tools provided by the GX\_Simulator graphical user interface (GUI). 
        
        However, the development, upgrade or maintenance of such tools is not easily sustainable in the absence of appropriate funding opportunities focused on supporting code development efforts in par with supporting the science that may - or may not - include some funding for proper code development and documentation.

    \subsubsection{Machine Learning-ready Data Sets}        
        Despite its ubiquitous use in the literature, rarely a definition is given for the term ``Machine Learning-Ready data''. Looking at the publicly available data sets claimed to be ML-ready, we can outline their common denominator as follows: (1) \textit{Accessibility}: the data set should be accessible to its designated users. Moreover, although a data set can in fact be commercial and therefore with limited access, in the scope of this white paper, it might be fair to assume that free and easy data accessibility to the entire community is implied and expected by the term ML-ready; (2) \textit{Completeness}: users must be given access to the entire collection, i.e., all attributes and all observations, and not only a subset of it; (3) \textit{Integrability}: all pieces of data that users may need for training, testing, and validation, and also for understanding their findings by linking the samples to the other relevant data sources must be available to the users without any extra effort or cost; (4) \textit{Sufficiency} (volume): the data set should be large enough for machine learning models. Deep learning algorithms are particularly known to be extremely data intensive. Therefore, if the problem of interest relies on such a family of algorithms, the data must be sufficiently large. While it is unknown how this size can be determined \citep[see a ``rough rule of thumnb'' in][pp. 20-21]{goodfellow2016deep}, factors such as the dimensionality of the data and the complexity of the ML algorithms (i.e., the number of parameters to be optimized) are some useful hints;
        (5) \textit{Cleanliness}; a ML-ready data set must be pre-processed in such a way that it prevents both ``garbage in, garbage out'', and imposing any data biases to ML algorithms \citep{keogh2003ucr, chu2016data}.
        It is worth mentioning that cleanliness of data should not be equated with the necessary pre-processing efforts such as normalization and imputation of the values. The pre-processing pipeline is decided and designed by the ML users and should not be forced by the data. This is an important part of the Machine Learning research and engineering, and not part of the data preparation process \citep{gudivada2017data}; (6) \textit{Understandability}: the necessary domain knowledge should be simplified and documented for the users so that ML experts can actually have a basic understanding of the normalities and abnormalities in the data. Of course, this is not to (and cannot) replace the need for domain expertise, but to help ML experts not overlook basic mistakes. 
        
    \subsubsection*{ML-ready Data Sets at GSU}        
        DMLab at GSU has recently released a ML-ready dataset called the Space Weather Analytics for Solar Flares (SWAN-SF). It is a collection of multivariate time series (MVTS) of active region properties comprising over 4,000 regions and spanning over 9 years of the SDO data products. It includes 51 flare-predictive parameters, and integrates over 10,000 flare reports. This is a benchmark dataset meaning that it serves the community as a testbed for their flare prediction efforts, allowing a fair and informative comparison. The five non-overlapping partitions of SWAN-SF help reproducibility of the efforts by limiting the arbitrary data splitting strategies. This dataset is openly accessible at Harvard Dataverse Repository \citep{mvts4swa2020dataverse}, as well as via a REST API on DMLab's server \citep{mvts4swa2020api}. A thorough discussion on the methods used for data collection, cleaning, and pre-processing of the solar active region and flare data is published \citep{angryk2020multivariate} and the entire project is made open-sourced as well \citep{mvts4swa2020bitbucketurl}. Several studies investigated different applications of SWAN-SF, such as time series profiling \citep{ma2019profiling}, outlier detection from spatiotemporal data \citep{cai2020local}, PIL detection \citep{cai2020framework}, synthetic MVTS generation \citep{chen2021towards}, robust sampling in extreme class-imbalance cases \citep{ahmadzadeh2021how}, flare classification using Time Series Forest \citep{ji2020all}, SEP forecasting \citep{ji2021modular}, and MVTS feature ranking \citep{atharv2021features}. Additionally, a data challenge series was organized by DMLab on this benchmark data set, as part of the IEEE Big Data Conference, in 2019 and 2020 \citep{dmlab2019datacup, dmlab2020datacup}.

    \subsubsection*{ML-ready Data Sets at NJIT}        
        The ML-ready data sets and associated ML source codes maintained by NJIT's Institute for Space Weather Sciences (ISWS) can be downloaded openly from the SolarDB cyberinfrastructure \citep{solardb} and also from the DeepSun site on GitHub \citep{deepsuncode}. Currently the SolarDB cyberinfrastructure contains three ML-ready data sets including a flare database, a global H$\alpha$ network, and a Big Bear Solar Observatory (BBSO) data archive. The flare database contains large flares (M5.0 or larger) in solar cycle 24, their active region information, and observations (solar images) from NASA's Solar Dynamics Observatory (SDO) that are collected 12 hours before and 6 hours after each large flare. The global H$\alpha$ network contains H$\alpha$ observations collected from eight observatories worldwide. The BBSO data archive contains historical observations collected by BBSO.
    
        In addition, SolarDB contains an operational near real-time flare forecasting system implemented based on a long short-term memory network and SHARP magnetic parameters \citep{Liu2019}. To make the work sustainable, it is important to pursue community involvement. We plan to extend the cyberinfrastructure to include ML-ready data sets and associated ML tools developed from other groups around the world, and also to categorize the data sets and tools based on their functions and capabilities. For example, one category may contain ML-ready data sets and ML tools for flare prediction and another category may contain ML-ready data sets and ML tools for Stokes inversion.

\section*{Acknowledgments}
This material is based upon work partly supported by the NSF 1743321, 1927578, 1639683 and NASA 80NSSC20K0302 grants to the New Jersey Institute of Technology, and by the NSF 1936361 and NASA 80NSSC22K0272 grants to the Georgia State University. The National Solar Observatory is operated by the Association of Universities for Research in Astronomy, Inc. (AURA) under cooperative agreement with the National Science Foundation. Any opinions, ﬁndings, and conclusions or recommendations expressed in this material are those of the authors and do not necessarily reﬂect the views of the funding agencies. 

\appendix
\section{Executive Summary of the 2014 Earth-Cube End-User Workshop: Science-Driven Cyberinfrastructure Needs in Solar-Terrestrial Research \label{STR}}
This section reproduces the Executive Summary of the 2014 Earth-Cube End-User Workshop {\it``Science-Driven Cyberinfrastructure Needs in Solar-Terrestrial Research,"} which was held at the New Jersey Institute of Technology, Newark, New Jersey, 2014 August 13-15.

The workshop was organized under the guidance of a Steering Committee composed by Gelu M. Nita (CSTR-NJIT), Dale E.  Gary (CSTR-NJIT), Andrew J. Gerrard (CSTR-NJIT), Gregory D. Fleishman (CSTR-NJIT), Alexander G. Kosovichev (CSTR-NJIT), Vincent Oria (CS-NJIT), and Marek Rusinkiewicz (CS-NJIT). 

\subsection{Introduction}
More than 80 domain scientists and students from three sub-disciplines of Geospace research (solar/heliospheric, magnetospheric, and upper-atmospheric research), as well as computer science, met at the Center for Solar-Terrestrial Research at New Jersey Institute of Technology for a 3-day workshop to examine the field’s current state of cyberinfrastructure (CI) and its future needs.  To prepare for the workshop, the steering committee identified 17 CI-knowledgeable leaders who represented each of the NSF Geosciences programs SHINE, GEM and CEDAR, as well as computer science.  This scientific organizing committee identified an additional 40+ scientists for invitation to the workshop, as well as NSF program managers Eva Zanzerkia (Earthcube), Ilia Roussev (SHINE), Anne-Marie Schmoltner (CEDAR), and Raymond Walker (GEM). 
The organizers endeavored to balance the demographics among the sub-disciplines and in relative experience of the participants.  Approximately 25\% of participants were early-career (8 students, 7 young scientists), 25\% mid-career, and 50\% in senior positions.  The sub-discipline participation was nearly evenly split, with 34\% SHINE, 23\% GEM, 23\% CEDAR, and 19\% computer science.  The preponderance of solar participation reflects mainly the concentration of solar research among the local NJIT participants.  The organizers believe that the workshop successfully captured the expertise and experience of the Geospace research community, and that the findings herein represent the consensus view of leaders and practitioners in science-driven cyberinfrastructure among space-science researchers. The full list of participants and their affiliations are listed in \S\ref{2014wokshop_participants}.

The Geospace disciplines are somewhat unique in the Geosciences for at least two reasons: (1) the disciplines are dominated by highly dynamic phenomena, and hence the data are organized mainly (though not entirely) on events and time rather than primarily spatially; and (2) the science drivers in these disciplines are studied in depth and decided upon as a broad-based community endeavor culminating in a decadal survey report every 10 years.  The most recent report, Solar and Space Physics: A Science for a Technological Society (National Research Council, The National Academies Press) was released in 2013, and serves as the main guide for science drivers examined during the workshop.  None of the findings below are meant to conflict in any way with the national science goals outlined in this decadal survey.

In addition to science goals, the NRC Decadal Survey also recommended, as a high priority, the implementation of an integrated initiative (DRIVE) to develop critical new technological capabilities in order to address the decadal survey’s complex scientific topics. In particular the decadal survey encourages the development of a ``data environment that draws together new and archived satellite and ground-based solar and space physics data sets and computational results from the research and operations communities." This included``community oversight of emerging, integrated data systems" and ``exploitation of emerging information technologies" with ``virtual observatories as a specific component of the solar and space physics research-supporting infrastructure."

\subsection{Science Issues and Challenges}
\subsubsection{Important science drivers}
The latest NRC Decadal Survey in Solar and Space Physics outlines four overarching key science goals for solar-terrestrial studies in the coming years. Below are more-focused science goals, consistent with the Decadal Survey goals, that we anticipate will benefit most from investments in cyberinfrastructure during the next 5 - 15 years:
\begin{itemize}
\item Understanding the couplings among physically different domains ranging from the solar interior to the Earth’s atmosphere: The advent of “Big Data” (the aggregation of large, complex, heterogeneous data sets) in observations and numerical modeling holds promise for rapid progress in solar-terrestrial research. Space- and ground-based observatories will provide important constraints for models in terms of boundary conditions and synthetic observables.  New observational data and computational advances provide new opportunities to develop cutting edge, data-driven models for the evolution of the magnetic flux below and above the solar surface, its influence throughout the heliosphere, and its impact at Earth. New cyberinfrastructure is required to improve our knowledge of the transfer of physical drivers across different physical domains from observational data and numerical simulations.
\item The study of the fundamental processes through which magnetic energy is generated, stored, released, and propagated: This is critically dependent on an advanced cyberinfrastructure that enhances our ability to assemble, analyze, and visualize multi-instrument, multi-wavelength data sets covering multiple temporal and spatial scales in combination with detailed physical models. The application of computer vision and machine learning techniques to identify features across different physical dimensions and to better mine large, distributed databases will be needed to enable event identification and statistically driven analysis.  Of particular interest is understanding the process of magnetic reconnection, the primary mechanism for energy release in solar flares and coronal mass ejections, which controls the occurrence and severity of magnetic storms through transport of mass, energy and momentum both at the sunward side of the magnetosphere and in the magnetotail.
\item Predicting the solar wind and Interplanetary Magnetic Field in the near-Earth environment. Understanding the origin of magnetic flux structure at the Sun, and how it evolves during magnetic eruption and propagation through the heliosphere to produce the relevant spatial scale of Bz variation near Earth that drives magnetic storms, will depend critically on in situ and remote sensing observations from the Solar Dynamics Observatory, Magnetospheric Multiscale, Solar Probe Plus and Solar Orbiter and other spacecraft, as well as ground-based facilities, combined with modeling techniques capable of simulating CME flux ropes from the Sun to the Earth.  The many disparate types of data and the broad range of spatial and temporal scales involved in both observations and models present a substantial cyberinfrastructure challenge.
\item Understanding the acceleration of particles throughout the Sun-Earth system.  Acceleration of electrons and ions, often to extremely high energies, is ubiquitous throughout the solar atmosphere, heliosphere, magnetosphere, and ionosphere, and creates hazards for humans and technological systems (spacecraft, communication and navigation systems, and even aircraft) everywhere within Geospace.  In every region, important tasks remain, such as: identifying the acceleration mechanisms that operate in the various regions of the Sun-Earth system;  determining which mechanisms are most important at different times and locations;  identifying common vs. distinct mechanisms in different regions; identifying the more important plasma instabilities that operate in the different regions and the role they play in particle acceleration under varying conditions; and following the propagation of accelerated particles within and across regions of the Sun-Earth system.
\item Understanding and forecasting the effects of forcing on the coupled Ionosphere-Thermosphere- Mesosphere (ITM) system.    The ITM system presents a unique challenge in that strong coupling between charged and neutral species dominates physical processes.   The system is responsive to external forces, e.g. reconnection, which impose global electric fields and magnetic currents, but also to internal processes, e.g. tropospheric heating and upward transmission of tidal forces, ionospheric instabilities, ion-neutral collisions and frictional drag.   The coupled system demands cross-disciplinary study involving data acquired over multiple time and distance scales from ground and space observatories.  Our ability to facilitate telecommunication and navigation, prevent catastrophic failure of the power grid during magnetic storms, or protect space assets from collisions demands accurate forecasting of the ITM response to forcing.  Unique to this effort, international collaborations often require the participation of poorer countries with desirable locations for observations, but without the means to install instrumentation or distribute data in optimal ways.
\end{itemize}

\subsection{Current Challenges to High-Impact, Interdisciplinary Science:}
The main challenges identified by workshop participants center around bridging the gaps among the Geospace sub-disciplines, to foster  interdisciplinary research.
\subsubsection{Challenges in finding / discovering data}
 \begin{itemize}
\item Users do not know how to search for data across multiple repositories, and in general what data sets/resources exist.  Data are hard to find, and even harder to transform into the form needed for further analysis.
\item Semantic techniques should be available to enable broad discovery and use of data.  Tools/libraries that enable the generation of metadata (annotations) in an automated fashion would be preferred.
\item Joint data discovery ideally makes use of of centralized data repositories  or search facilities where all the metadata (and pointers to the data) are queried and made available through a common interface. Complementary to this would be the implementation of  a system based on semantic web technologies, which would require that a widely accepted standard vocabulary/ontology (suitable for our community) be put in place that the community agrees to abide to.
\item There is a need for encouraging adoption and consistent usage of metadata standards for the essential attributes of both observational and modeling data sets, as well as an agreement on vocabulary to use.
\item Getting to a set of “widely accepted standards” is itself a challenge.  Also needed are translation tools (“ontology alignment”) between different sets of standards, especially where there are already multiple sets of established practices.
\item The Geospace disciplines increasingly need better tools for mining our spatiotemporal data sets for features, both known and unknown
\item The tools need to be scalable, to work for both large and small data sets.
\item Data query: enabling the easy and effective querying of very specific subsets of data in order to tailor the results according to a specific science objective, thus reducing the volume of the data transfer. Good metadata and strong quick-look tools play a big role in this.
\item Data volumes are becoming prohibitively large.  It is not feasible to co-locate all data sets, or even apply the “old model“ of requiring users to download all the data sets of interest onto their own computers to manipulate them locally.  Analysis increasingly needs to be co-located with the data, but this is problematic for analysis of multiple data sets, located in different places. Processing and user-driven analysis carried out at these large data centers may provide a solution to this coming problem, but mechanisms need to be in place to allow these providers to develop and support these (potentially costly) capabilities. 
\end{itemize}
\subsubsection{Challenges in working with data}
 \begin{itemize}
\item Continuity of data sets (both space and ground-based) over time has an increasing value as our ability to mine and probe these large data collections grows.  Ensuring continuity should be a factor in funding decisions. (For example, there are concerns about several older instruments with no successor at the moment.)
\item There are similar issues of continuity in the development of data analysis tools as well as instruments.
\item Getting the most out of existing or legacy data; ensuring things do not get lost over time as missions or groups end.
\item Information about assumptions, sources of error, and methodologies should be included along with the data.
\item Need methods to ensure scientific reproducibility by allowing citation of specific data products and processing steps used in a scientific study.
\item Need a mechanism for ensuring proper attribution of data sources in publications.  It is critical to record provenance of all data to improve future reuse.
\item Need better benchmarking/validation of data catalogs for researchers in different disciplines: it is important to have clear quality metrics that allow users to determine which data points are “good” or “bad” for their purposes.
\item It is important not to “re-invent the wheel.”  If someone has “solved” a problem, other communities need to be able to find out about this and make use of it.
\item The wide variety of analysis tools and languages in current use inhibits the development of a common set of analysis tools. Clearer documentation and use of software development best practices would help mitigate this confusion.
\item There is a need for a strong leadership structure: a project should be run by a single, strong entity with broad community buy-in to ensure coordination.
\end{itemize}
 \subsubsection{Challenges in cross-disciplinary science / working with data outside our sub-discipline}
 \begin{itemize}
\item Data from outside a researcher’s field is difficult to find and learn how to analyze.
\item An impediment to cross-disciplinary research is that while the same problems might be studied in different sub-disciplines, the observables, scales, and parameter regimes may be quite different.
\item It is difficult to find sources of funding for cross-disciplinary research.
\item Researchers using data from outside their areas of expertise need trusted catalogs of events and categorizations
\item Data integration is needed to enable interfacing and interoperability among diverse data sets.
\item Need better support for ‘sun-to-mud’ efforts.  Solutions may be to have more common workshops, and classes offered online by multiple institutions.
\end{itemize}
 \subsubsection{Modeling-specific challenges}
 \begin{itemize}
\item It is important to compare and address discrepancies between data and models.  Tools are generally not readily available to directly compare model outputs and observations.
\item If these tools were available, iteration between modeling and data comparison could take place, allowing ongoing improvement of both.
\item While data are often open and analysis code is sometimes open source, the same is not generally true for models (although it should be).
\item  In terms of modeling: there is a need for better flexibility/modularity in large model design so various groups could “plug and play” their components.
\end{itemize}
\subsubsection{Educational, societal, and public outreach challenges}
 \begin{itemize}
\item There is a dearth of data-science and cyberinfrastructure-related content in the domain-specific academic curricula, impairing the ability of students to incorporate existing tools and best practices into their research.
\item Scientists often do not know how to scale up their cyberinfrastructure usage from the desktop to make use of high-performance computing (HPC).
\item Students and practicing researchers need training on how to use GPUs and other advanced computing resources.
\item Scientists want to share their data in the public domain, but may worry about potential misuse or misinterpretation of the data.
\end{itemize}
\subsection{Technical Issues/Challenges}
Many of the interdisciplinary science challenges noted above are rooted in technical issues that must be addressed in order to successfully overcome them.  The breakout sessions devoted to technical challenges included moderators who are computer scientists, in order to encourage new thinking.
\begin{itemize}
\item There is a need to develop computationally efficient capabilities for searching and expressive querying of Large/Diverse/Distributed Data Sets including provenance and data quality. What is of interest to scientists can be very complex to define.  With today’s high-volume databases, it is increasingly important to locate and download only the portion of data of interest.  Propagation delays from one regime to another within the Geospace system make event searches challenging—e.g. how to do correlations to find linked events among data sets with such delays, without downloading all of the data.
\item There will be a continuing need to discover, search, and utilize historical data sets, which must be preserved and, if necessary, modernized through metadata indexing to bring them into discoverable form.
\item Data providers, especially new and actively maintained services, need to include well-documented APIs (application programming interfaces) and service interfaces, to aid in development of flexible workflows for utilizing the data resources.
\item Some metadata standards already exist, but translators/converters are needed for searches bridging solar-terrestrial environments (solar, heliosphere, magnetosphere, ionosphere/upper-atmosphere) to promote interdisciplinary science.  Additional efforts to agree on a wider standard of keywords, vocabulary and ontologies would be useful, but difficult.
\item A platform and standards for data and software citations need to be further developed and widely adopted. A scheme for searching ranked databases and software according to popularity, usage, and quality would be a useful addition.
\item Workshops/tutorials and academic curricula are needed to teach standard tools and techniques for interdisciplinary research  to the community (e.g., orbital discovery tool). Community-developed toolkits (e.g. those at SolarSoft, sunpy.org, itk.org) are important sources of cross-platform tools for general analysis.  Community involvement in further open-source tool development (e.g. through Github) should be strengthened and encouraged.
\item Tools are needed for generalized Event/Object recognition in space and time, and for visualizing multi-dimensional data in large data volumes
\end{itemize}
\subsection{Community Next Steps}
Since this Solar-Terrestrial Cyberinfrastructure workshop occurred rather late in the process of Earthcube governance, we have the advantage of knowing the context of the program within which we should coordinate our efforts.  Many of the challenges identified during the workshop have also been identified by other domain workshops, and hence our community can form Earthcube working groups or join with others already forming within Earthcube.  In addition, our community can undertake the following steps, and also encourage NSF to provide Earthcube funding opportunities to address these areas:
\subsubsection{Tools and Standards}
\begin{itemize}
\item Make/collect a list of useful tools and services (with user reviews)
\item Provide additional tools for generating metadata  from existing data and manipulating metadata in the form of plots, indexing
\item Support development of community-led general analysis toolkits
\item Provide translators between standard data formats
\item Provide translators between metadata (e.g. keywords) standards
\item Develop standard service interfaces (such as APIs)
\item Develop “one-stop-shopping facility” to aggregate data, or facilitate ordering/delivery of data
\end{itemize}
\subsubsection{Cross-disciplinary CS/domain scientists collaborations} 
\begin{itemize}
\item Assemble domain scientists and computer scientists to attack specific and realistically achievable high-value science goals as identified by the decadal survey
\item Identify and list the most widely-used data-sets in the relevant disciplines and design data integration tools according to the above-mentioned science goals
\item Create hyper-dimensional visualization tools
\item Develop the capability for advanced semantic queries for nearest-neighbor matching of widely dissimilar data
\item Develop the capability to construct queries of what is missing (identifying gaps and dealing with intermittency in data coverage)
\end{itemize}
\subsubsection{Education (community and academic)}
\begin{itemize}
\item Adding cyberinfrastructure and computer visualization components to solar-terrestrial curricula.
\item Educating domain scientists on scaling up their applications from desktop to HPC
\item Access to HPC resources for training in solar-terrestrial research
\item Education on how to utilize GPU and other advanced computing resources
\item Advanced data analysis techniques (e.g. inverse theory, forward fitting, data assimilation)
\end{itemize}
\subsubsection{Data management}
\begin{itemize}
\item Searching and querying long-term archived databases with access control and provenance
\item Use of DOIs and alternatives for data and software citations
\item Tools and standards for creation of metadata that tracks database use (who, for what purpose, popularity)
\item Cloud storage and HPC processing
\item Support for creation, population, and operation of new databases based on new instruments and modeling efforts
\item Capability for creation of quick-look data products
\end{itemize}
\subsubsection{Model input/output}
\begin{itemize}
\item Develop techniques for data-assimilation, data-driven modeling, and cross-domain model coupling
\item Metadata concepts for model output (descriptive of format)
\item Develop standards and guidelines for making model output shareable and comparable
\item Search tools for integrating observational and model output data
\end{itemize}
\subsubsection{Quantifying data quality}
\begin{itemize}
\item Include valid error estimates together with data
\item Include information about data quality, completeness, and fitness for use
\item Research methods and practices for quantifying errors (random, systematic)
\item Biases introduced by data processing
\end{itemize}
\subsubsection{Encouraging good practices}
\begin{itemize}
\item Study feasibility of creating cloud-storage for data, whose use would enforce good practices as a prerequisite for use
\item Create or join an Earthcube working group to identify and share information and tools for enforcing metadata standards
\item Include software engineering and development techniques as part of academic training
\end{itemize}
\subsection{Workshop Participants \label{2014wokshop_participants}}
This workshop was attended by 80 participants representing 41 institutions, as listed below.

\subsubsection{List of Participants}

Nabil Adam$^{ 1}$, Rafal Angryk$^{ 2}$, Kemafor Anyanwu$^{ 3}$, Tim Bastian$^{ 4}$, Jacob Bortnik$^{ 5}$, William Bristow$^{ 6}$, Gary Bust$^{ 7}$, Rebecca Centeno$^{ 8}$, Andrew Cerbone$^{ 9}$, Mark Cheung$^{ 10}$, Peter Chi$^{ 5}$, Ross Cohen$^{ 9}$, Russell Cosgrove$^{ 11}$, Alfred de Wijn$^{ 8}$, Darren De Zeeuw$^{ 12}$, William Denig$^{ 13}$, Eric Donovan$^{ 14}$, Alexander Engell$^{ 15}$, Tzu-Wei Fang$^{ 16}$, Gregory Fleishman$^{ 9}$, Dale Gary$^{ 9}$, Hongya Ge$^{ 9}$, Andrew Gerrard$^{ 9}$, Jesper Gjerloev$^{ 7}$, Lindsay Glesener$^{ 17}$, Dalkandura Arachchige Kalpa Gunaratna$^{18}$, Joseph Gurman$^{ 19}$, Frank Hill$^{ 20}$, Cheryl Huang$^{ 21}$, Jack Ireland$^{22}$, Ramesh Jain$^{ 23}$, Sarah Jones$^{ 19}$, Hyomin Kim$^{ 24}$, Irina Kitiashvili$^{ 25}$, Craig Kletzing$^{ 26}$, Alexander Kosovichev$^{ 9}$, Kate Kretschmann$^{ 27}$, Louis J. Lanzerotti$^{ 9}$, Marc Lessard$^{ 28}$, Hanli Liu$^{ 29}$, Gang Lu$^{ 29}$, Xiguo Ma$^{ 9}$, Peter MacNeice$^{ 19}$, Marlo Maddox$^{ 19}$, Jonathan Makela$^{ 30}$, Nagi Mansour$^{ 31}$, Jerry Manweiler$^{32}$, Dhvanit Mehta$^{ 9}$, Ethan Miller$^{ 7}$, Viswanath Nandigam$^{ 33}$, Kuroda Natsuha$^{ 9}$, Gelu Nita$^{ 9}$, Vincent Oria$^{ 9}$, Nicholas Pedatella$^{ 34}$, David Perel$^{ 9}$, Kevin Reardon$^{ 20}$, William Rideout$^{ 35}$, Ilia Roussev$^{ 36}$, Marek Rusinkiewicz$^{ 9}$, Michelle Salzano$^{ 9}$, Ludger Scherliess$^{ 37}$, Anne-Marie Schmoltner$^{ 36}$, Shaheda Shaik$^{ 9}$, Amit Sheth$^{ 18}$, Atreyee Sinha$^{ 9}$, Andrei Sirenko$^{ 9}$, Jichao Sun$^{ 9}$, Krishnaprasad Thirunarayan$^{ 38}$, Kevin Urban$^{ 9}$, Anthony van Eyken$^{ 11}$, Rodney Viereck$^{ 39}$, Raymond Walker$^{ 5}$, Zhitao Wang$^{ 9}$, Haimin Wang$^{ 9}$, Allan Weatherwax$^{ 40}$, Robert Weigel$^{ 41}$, Michael Wiltberger$^{ 29}$, Yan Xu$^{ 9}$, Vasyl Yurchyshyn$^{ 9}$, Eva Zanzerkia$^{ 36}$

\subsubsection{Affliations}
$^{ 1}$Rutgers University Newark, $^{ 2}$Georgia State University, $^{ 3}$North Carolina State University, $^{ 4}$NRAO, $^{ 5}$UCLA, $^{ 6}$University of Alaska Fairbanks, $^{ 7}$JHUAPL, $^{ 8}$HAO, $^{ 9}$NJIT, $^{ 10}$Lockheed Martin Solar \& Astrophysics Laboratory, $^{ 11}$SRI International, $^{ 12}$University of Michigan, $^{ 13}$National Geophysical Data Center, $^{ 14}$University of Calgary, $^{ 15}$Flare Forecast, $^{ 16}$CU/CIRES, $^{ 17}$UC Berkeley, $^{ 18}$Kno.e.sis Center, $^{ 19}$NASA GSFC, $^{ 20}$NSO, $^{ 21}$Air Force Research Laboratory, $^{22}$ADNET Systems, $^{ 23}$UC Irvine, $^{ 24}$Virginia Tech, $^{ 25}$NASA Ames Research Center, $^{ 26}$University of Iowa, $^{ 27}$Arizona Geological Survey, $^{ 28}$Univ of New Hampshire, $^{ 29}$NCAR, $^{ 30}$University of Illinois at Urbana-Champaign, $^{ 31}$NASA Ames Research Center, $^{32}$Fundamental Technologies, $^{ 33}$University of California San Diego, $^{ 34}$UCAR / COSMIC, $^{ 35}$MIT Haystack, $^{ 36}$NSF, $^{ 37}$Utah State University, $^{ 38}$Wright State University,$^{ 39}$NOAA/NWS/Space Weather Prediction Center, $^{ 40}$Siena College, $^{ 41}$George Mason University 
\bibliography{ref.bib}

\end{document}